\documentclass[amsmath,reprint,showkeys,showpacs,nofootinbib,longbibliography]{revtex4-1}
\usepackage{amsfonts,amssymb}
\usepackage{graphicx} 
\usepackage{bbm}
\usepackage{bm}

\PassOptionsToPackage{unicode}{hyperref}
\PassOptionsToPackage{naturalnames}{hyperref}
\def\beq{\begin{equation}}
\def\eeq{\end{equation}}
\def\bsubs{\begin{subequations}}
\def\esubs{\end{subequations}}
\def\bal#1\eal{\begin{align}#1\end{align}}

\usepackage[hypertexnames=false]{hyperref}
\begin{document}
\title{Post--selection induced deterministic and probabilistic entanglement with strong and weak interactions}
\author{Antonio Di Lorenzo}
\affiliation{Instituto de F\'{\i}sica, Universidade Federal de Uberl\^{a}ndia, Av. Jo\~{a}o Naves de \'{A}vila 2121, 
Uberl\^{a}ndia, Minas Gerais, 38408-100,  Brazil}%
\begin{abstract}
A scheme is proposed to entangle two systems that have not interacted by using an ancillary particle in a Mach-Zehnder interferometer, by making a suitable post--selection of the particle followed by a conditional feedback on one of the subsystems to be entangled. 
For a strong interaction, the process works deterministically. For a weaker interaction only the probability of success is reduced, but the output continues to be a maximally entangled state. 
\end{abstract}
\maketitle
It is essential to produce entangled states on demand, both for quantum computation and for quantum cryptography. 
Furthermore, besides its potential for applications, entanglement is by itself interesting on a fundamental level\cite{Einstein1935,Schrodinger1935,Bell1964}. 
A challenging problem is to produce entangled particles deterministically, or at least with high efficiency. Here, we show that it is possible to do so by using a single particle as an ancilla, and 
by making a measurement--based local feedback.  

The setup proposed in this Letter has its roots in the 90s, when, following a proposal by Hardy \cite{Hardy1994} , later clarified by Peres \cite{Peres1995}, Gerry proposed an experiment using cavity quantum electrodynamics to demonstrate the ``non-locality'' of a single photon \cite{Gerry1996}, i.e. that a single photon can form an entangled state with the vacuum, a phenomenon called also single--photon entanglement. 
Gerry's proposal was further elaborated by Moussa and Baseia \cite{Moussa1998}. 
In later years, some controversy arose on the subject whether the entanglement between a photon and the vacuum is factual or merely formal \cite{Pawlowski2006,Enk2005,Drezet2006,Enk2006}. Today there are several experimental realizations of entanglement swapping and teleportation using single--photon entanglement 
\cite{Sciarrino2002,Lombardi2002}.
More recently, Aharonov \emph{et al.} proposed a way to apparently separate a particle from its properties \cite{Aharonov2013}, a phenomenon dubbed the quantum Cheshire cat. An experimental realization was made \cite{Denkmayr2014}, even though it does not translate faithfully the prescription of the original proposal. Furthermore, the existence of the phenomenon relies on a controversial interpretation of the weak value \cite{DiLorenzo2014c}. In a previous paper \cite{DiLorenzo2014b}, we pointed out a connection between the Quantum Cheshire cat and the phenomenon of entanglement swapping from a single--photon entangled state to separate systems. Here, we consider a deterministic way of realizing said swapping, i.e. we show how entangled pairs can be created on demand if the interaction with the ancilla is strong, we point out how the procedure works also, but only probabilistically, if the interaction is weak. In the latter case, perhaps surprisingly, we show that maximally entangled states can be still extracted, however with a reduced probability of success. We also show how the environment and the internal degrees of freedom of the ancilla can degrade the performance, and how it is important that the ancilla couple with the same strength to each 
subsystem to be entangled. 

%

%

\begin{figure}[h!]
\centering
\includegraphics[width=3in]{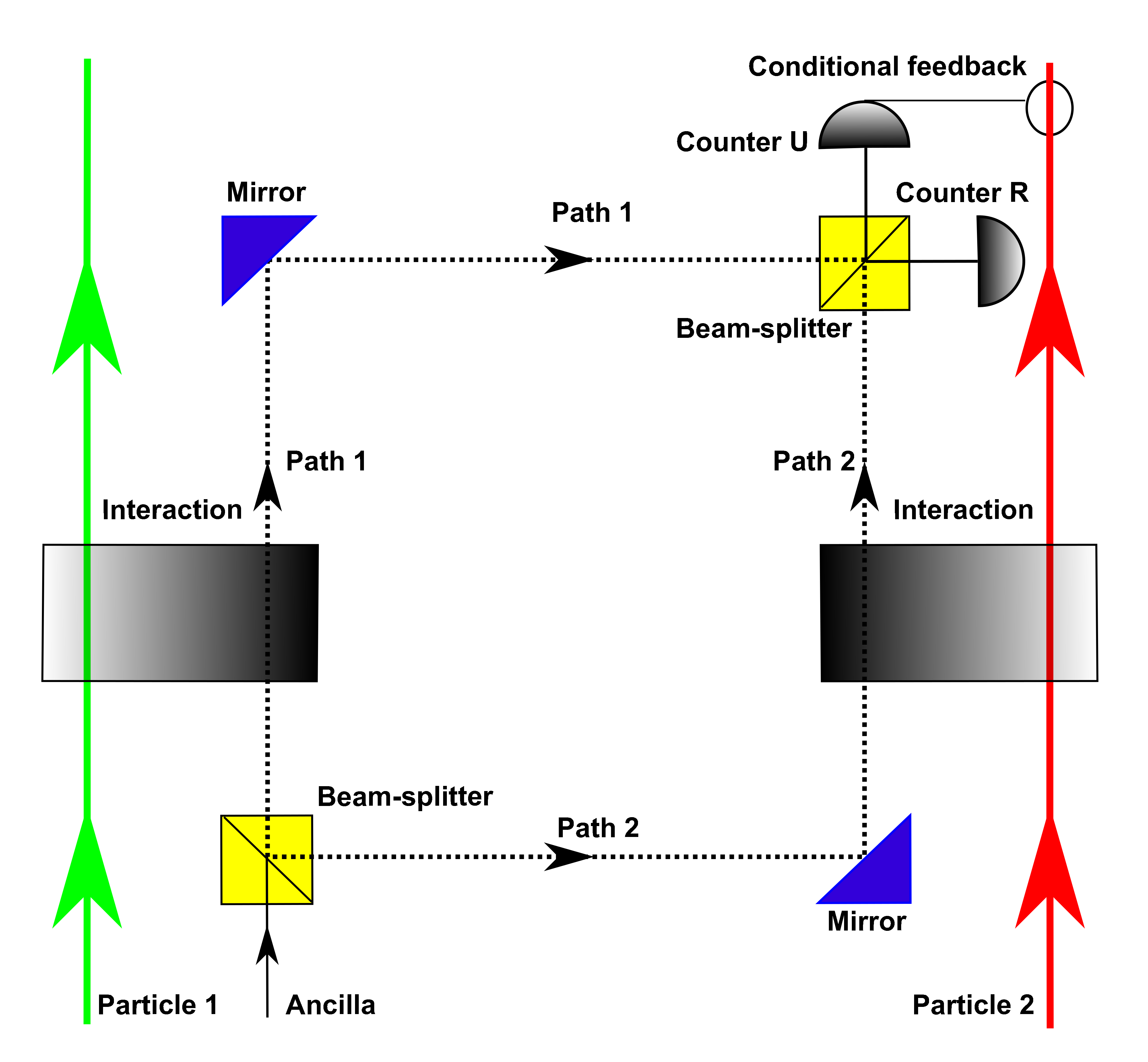}
\caption{\label{fig:setup} The setup is a Mach-Zehnder interferometer, where an ancillary particle interacts with the subsystems to be entangled, each placed at a different arm. 
If the particle counter $R$ clicks, the system $1+2$ is maximally entangled, irrespectively of the strength of the interaction (the latter being represented by a black box). 
If the particle counter $U$ clicks, the system $1+2$ is only partially entangled, unless the interaction is strong. In the latter case, by making an appropriate local operation, say, on the subsystem 2, the pair can be brought to the same maximally entangled state as in the case when $R$ clicks.}
\end{figure}
%
With reference to Fig.~\ref{fig:setup}, we have an ancillary particle that enters a Mach-Zehnder interferometer. If it takes path 1, the ancilla interacts with particle 1, and if it takes path 2, it interacts with particle 2. Before the ancilla enters the Mach-Zehnder interferometer, the initial preparation $|A_1,A_2,\Psi_{anc}\rangle$ 
for the system composed by the subsystems to be entangled (1 and 2) and by the ancillary particle (which could be, but need not to, a photon) is a pure factorable state. 
In a nondemolition measurement (see von Neumann's book \cite{vonNeumann1932}), 
if the ancilla, after the first beam splitter, follows path 1, and hence its state is described by $|\Psi_{anc}\rangle=|1,\mu\rangle$, with $\mu$ internal degrees of freedom, 
the total state, after the interaction represented by the grayed box, evolves to 
$|B_1,A_2,\{1,\mu\}\rangle$. If the ancilla follows path 2, instead, the final state after the interaction is $|A_1,B_2,\{2,\mu\}\rangle$. 
The states $|A_j\rangle$ and $|B_j\rangle$ are orthogonal if the interaction is strong, so that each subsystem would be performing a projective measurement of the presence 
of the ancilla, if it was observed in the basis spanned by $|A_j\rangle$ and $|B_j\rangle$. In general, however, the latter states are not orthogonal. For instance, this happens for a weak measurement, when  $|A_j\rangle$ and $|B_j\rangle$ are almost indistinguishable. 
Here, we shall not make any assumption about the strength of the measurement. 
Instead, we shall only exploit the fact, that, in general 
\beq
|B_j\rangle = \cos{\theta_j}|A_j\rangle + \sin{\theta_j} |A_j^\perp\rangle ,
\eeq
where $|A_j^\perp\rangle$ is orthogonal to $|A_j\rangle$. 
Without loss of generality, we can fix the phase in such a way that $0\le\theta_j\le\pi/2$. 
For $\theta_j\to \pi/2$ we have a strong interaction, for $\theta_j\to 0$ the interaction is weak. 

Because of the superposition principle, the state of the system composed by the particles 1 and 2, and by the ancilla, before the latter reaches the second beamsplitter, is   
\beq 
|\Psi_{tot}\rangle=\frac{1}{\sqrt{2}}\left[|B_1,A_2,\{1,\mu\}\rangle + |A_1,B_2,\{2,\mu\}\rangle \right].
\eeq
The reduced state of $1+2$, if the right particle--counter or the upper counter clicks is obtained by projecting the total state of $1+2+ancilla$ on the subspace 
represented, respectively, by the states $|R\rangle = (|1\rangle-|2\rangle)/\sqrt{2}$ and $|U\rangle = (|1\rangle+|2\rangle)/\sqrt{2}$ , namely
\begin{widetext}
\bsubs
\bal
&|\Psi_{1,2}\rangle_R = \frac{1}{2} \left[|A_1,B_2\rangle-|B_1,A_2\rangle\right]
=  \frac{1}{2} \left[\left(\cos{\theta_2}-\cos{\theta_1}\right)|A_1,A_2\rangle+\sin{\theta_2}|A_1,A_2^\perp\rangle-\sin{\theta_1}|A_1^\perp,A_2\rangle\right]
.
\label{eq:rightclick}
\\
&|\Psi_{1,2}\rangle_U = \frac{1}{2} \left[|A_1,B_2\rangle+|B_1,A_2\rangle\right] 
=  \frac{1}{2} \left[\left(\cos{\theta_2}+\cos{\theta_1}\right)|A_1,A_2\rangle+\sin{\theta_2}|A_1,A_2^\perp\rangle+\sin{\theta_1}|A_1^\perp,A_2\rangle\right]
.
\label{eq:upclick}
\eal
\esubs
\end{widetext}
The optimal result is obtained assuming an equal interaction strength $\theta_1=\theta_2=\theta$, which we shall assume in the first part of this Letter. 
In this case, indeed, the state \eqref{eq:rightclick} is a maximally entangled state, independently of the strength of the interaction, 
\beq
|\Psi_{1,2}\rangle_R 
=  \frac{\sin{\theta}}{2} \left[|A_1,A_2^\perp\rangle-|A_1^\perp,A_2\rangle\right]
.
\label{eq:rightclick2}
\eeq
The normalization was chosen in such a way that $|\Psi_{1,2}|^2$ equals the probability of post--selection, namely 
\bsubs
\bal
\langle \Psi|\Psi\rangle_R =&\ \frac{(\sin{\theta})^2}{2},
\\
\langle \Psi|\Psi\rangle_U =&\ \frac{1+(\cos{\theta})^2}{2}.
\eal
\esubs
Therefore, we reach an important conclusion: The post--selected state, when the right counter clicks, is always maximally entangled, independently of the strength of the interaction. 
For a weak interaction, the probability of success, however, tends to 0. 

On the other hand, for a strong interaction $\theta=\pi/2$, the state \eqref{eq:upclick} is also maximally entangled, 
\beq
|\Psi_{1,2}\rangle_U 
=  \frac{1}{2} \left[|A_1,A_2^\perp\rangle+|A_1^\perp,A_2\rangle\right]
.
\label{eq:upperclick2}
\eeq
Therefore, it can be reduced to the state 
\eqref{eq:rightclick2} by means of a local unitary transformation. 
We reach thus the second important conclusion: 
For a strong interaction, the two subsystems 1 and 2 can be entangled deterministically, using a post--selection based local feedback on either subsystem. 

Next, we consider the effect of the environment. We shall not make a specific dynamical model, which would depend on the physical realization of the proposed setup. 
Instead, we shall make a general, simplified treatment, inspired to the von Neumann approach to the measurement. 
Before the particle reaches either counter, the final state of the overall system formed by the subsystems $1$ and $2$, the ancilla, and the environment, 
is 
\beq
|\Psi_{tot}\rangle = 
|B_1,A_2,\{1,\mu\},\mathcal{E}_1\rangle +|A_1,B_2,\{2,\mu\},\mathcal{E}_2\rangle. 
\eeq
Here $|\mathcal{E}\rangle$ denote states of the environment. Depending which path the ancilla takes, the environment is affected differently. 
The final state of the system $1+2$ is obtained by projecting over the postselection state $|U\rangle=(|1\rangle+|2\rangle)/\sqrt{2}$ or 
$|R\rangle=(|1\rangle-|2\rangle)/\sqrt{2}$, and by tracing out the environment. In general, this procedure yields a mixed state, 
\begin{widetext}
\bal
&\rho_{1,2} = \frac{1}{4} \left[|A_1,B_2\rangle\langle A_1,B_2| + |B_1,A_2\rangle\langle B_1,A_2| 
\mp \gamma  |A_1,B_2\rangle\langle B_1,A_2|\mp\gamma^*  |B_1,A_2\rangle\langle A_1,B_2|\right]
,
\label{eq:envclick}
\eal
\end{widetext}
where the upper sign refers to the $R$ counter clicking and the lower sign to the $U$ counter clicking, while 
\beq
\gamma = \langle \mathcal{E}_1|\mathcal{E}_2\rangle
\eeq
is the overlap between the two different states of the environment. 
Even for a strong interaction, when $|A_j\rangle$ and $|B_j\rangle$ are orthogonal, the environment reduces the amount of entanglement. 
If the states $|\mathcal{E}_1\rangle$ and $|\mathcal{E}_2\rangle$ are perfectly distinguishable, i.e. $\gamma=0$, the entanglement is cancelled. 
The reason is that in this limit the environment works as a which--path detector, destroying coherence. 
However, if we are able to observe the states of the environment, which means that we are not tracing out its degrees of freedom, we have still an entangled state, 
$|B_1,A_2,\mathcal{E}_1\rangle-|A_1,B_2,\mathcal{E}_2\rangle$.  
We remark that the internal degrees of freedom of the ancilla may work effectively as an environment, if they change to different internal states on different paths, 
$|\{1,\mu\}\rangle\to |\{1,\mu_1\}\rangle$ and $|\{2,\mu\}\rangle\to |\{2,\mu_2\}\rangle$. 

\begin{figure}
\centering
\includegraphics[width=3in]{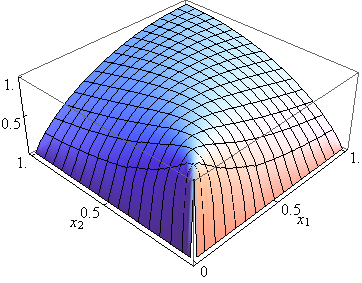}
\caption{\label{fig:ent} The entropy of entanglement, normalized to one, for the state given in Eq.~\eqref{eq:rightclick}, as a function of $x_1=1-\cos{\theta_1}$ and $x_2=1-\cos{\theta_2}$.}
\end{figure}

Finally, we consider the more general case, when the interaction strength differs in each arm. 
In Fig.~\ref{fig:ent} we plot the entropy of entanglement of the state \eqref{eq:rightclick}
as a function of $x_1=1-\cos{\theta_1}$ and $x_2=1-\cos{\theta_2}$, which can be considered effective, dimensionless coupling strengths. 
The analytic formula is easily found to be 
\bal
S =&\ -
\frac{x_1\left(1-\frac{x_2}{2}\right)}{x_1+x_2-x_1x_2}\log_2\left(\frac{x_1\left(1-\frac{x_2}{2}\right)}{x_1+x_2-x_1x_2}\right)
\nonumber
\\
&- \frac{x_2\left(1-\frac{x_1}{2}\right)}{x_1+x_2-x_1x_2}\log_2\left(\frac{x_2\left(1-\frac{x_1}{2}\right)}{x_1+x_2-x_1x_2}\right).
\eal
In the weak measurement regime, the entanglement is very unstable against small differences in coupling strengths. Strongest couplings are therefore to prefer, not only 
because they allow to produce entanglement on-demand, but also because the entanglement is robust against small asymmetries in the couplings.

In conclusion, we have demonstrated a technique to entangle two quantum systems, by having them interact with an ancillary particle in a Mach-Zehnder interferometer and then by post--selecting the ancilla. We have shown that if the interaction is strong, the entanglement is obtained with 100\%  efficiency by making an appropriate feedback, while, if the interaction is weak, the fraction of the systems corresponding to a ``wrong'' post--selection should be discarded in order to obtain a maximally entangled state. 
In perspective, there is an interesting optimization problem related to this proposal. Let $G$ the gain associated to a unit of entanglement and let $L$ the loss due to discarding a pair when the $U$ detector clicks. We should find the optimal fraction $w$ of pairs to retain, which would be otherwise discarded, and the optimal local unitary transformation $\mathcal{U}$, such that the net gain 
\bal
N =&\ \frac{G}{2}\left[\sin^2\theta+w(1+\cos^2\theta)\right]E_1[\rho_{1,2}(w,\mathcal{U})] 
\nonumber
\\
&- \frac{L}{2} (1-w) (1+\cos^2\theta) 
\eal  
is maximized.  
Here $E_1$ is a measure of entanglement with maximum value 1, and $\rho_{1,2}(w,\mathcal{U})\propto|\Psi_{1,2}\rangle\langle \Psi_{1,2}|_R + w \mathcal{U}|\Psi_{1,2}\rangle\langle \Psi_{1,2}|_U\mathcal{U}^\dagger$ is the mixed state obtaining by picking all the occurrences when $R$ clicks and a fraction $w$ of the occurrences when $U$ clicks, in which case the local unitary $\mathcal{U}$ is applied. 

During the completion of the present work, Ohm and Hassler have proposed a realization of the deterministic scheme with strong measurement by using a photon as an ancilla and transmon qubits as systems 1 and 2 \cite{Ohm2015}. Other possibilities of experimental realizations include photons interacting through a non--linear Kerr medium and atoms in two cavities. 

\begin{acknowledgments}
This work was performed as part of the Brazilian Instituto Nacional de Ci\^{e}ncia e
Tecnologia para a Informa\c{c}\~{a}o Qu\^{a}ntica (INCT--IQ), it
was supported by Funda\c{c}\~{a}o de Amparo \`{a} Pesquisa do 
Estado de Minas Gerais through Process No. APQ--01863--14 and 
by the Conselho Nacional de Desenvolvimento Cient\'{\i}fico e Tecnol\'{o}gico (CNPq) 
through Process No. 311288/2014--6. 
\end{acknowledgments}
\bibliography{../weakmeasbiblio}
\end{document}